\documentclass{emulateapj}

\newcommand\vect[1]{\mbox{\boldmath{$#1$}}}
\newcommand\tjptensor[1]{\mbox{{$\bf #1$}}}
\newcommand{\Kel}{\,{\rm K} }
\newcommand{\Jy}{\,{\rm Jy} }
\newcommand{\mJy}{\,{\rm mJy} }
\newcommand{\muKt}{\,\mu{\rm K}^2 }

\begin{document}

\journalinfo{Accepted for publication in The Astrophysical Journal}
\submitted{}
\shortauthors{MASON ET AL.}
\shorttitle{DEEP FIELD OBSERVATIONS WITH THE COSMIC BACKGROUND IMAGER}

\title{The Anisotropy of the Microwave Background to $\ell = 3500$:
Deep Field Observations with the Cosmic Background Imager}

\author{B. S. Mason,\altaffilmark{1} T. J. Pearson, A. C. S. Readhead,
M. C. Shepherd, J. Sievers, P. S. Udomprasert, J. K. Cartwright,
A.J. Farmer, and S. Padin} \affil{Owens Valley Radio Observatory,
California Institute of Technology, 1200 East California Boulevard,
Pasadena, CA 91125} \author{S. T. Myers} \affil{National Radio
Astronomy Observatory, P.O. Box O, Socorro, NM 87801}
\author{J. R. Bond, C. R. Contaldi, U. Pen, and S. Prunet}
\affil{Canadian Institute of Theoretical Astrophysics, 60 St. George
Street, Toronto, Ontario M5S 3H8} \author{D. Pogosyan} \affil{Avadh
Bhatia Physics Laboratory, University of Alberta, Edmonton, AB T6G 2J1}
\author{J. E. Carlstrom, J. Kovac,  E. M. Leitch, and C. Pryke}
\affil{University of Chicago, 5640 S. Ellis Ave., Chicago, IL 60637}
\author{N. W. Halverson, W. L. Holzapfel} \affil{University of California, 426 Le Conte
Hall, Berkeley, CA 94720-7300} \author{P. Altamirano, L. Bronfman,
S. Casassus and J. May} \affil{Departamento de Astronom\'{\i}a,
Universidad de Chile, Casilla 36-D, Santiago, Chile} \and
\author{M. Joy} \affil{Dept. of Space Science, SD50, NASA/Marshall
Space Flight Center, Huntsville, AL 35812}

\altaffiltext{1}{Current Address: National Radio Astronomy
  Observatory, P.O. Box 2, Green Bank, WV 24944}

\begin{abstract}

We report measurements of anisotropy in the cosmic microwave
background radiation over the multipole range $\ell \sim 200
\rightarrow 3500$ with the Cosmic Background Imager based on deep
observations of three fields. These results confirm the drop in power
with increasing $\ell$ first reported in earlier measurements with
this instrument, and extend the observations of this decline in power
out to $\ell \sim 2000$.  The decline in power is consistent with the
predicted damping of primary anisotropies.  At larger
multipoles, $\ell = 2000$--$3500$, the power is $3.1\sigma$ greater
than standard models for intrinsic microwave background anisotropy in
this multipole range, and $3.5 \sigma$ greater than zero.  This excess
power is not consistent with expected levels of residual radio source
contamination but, for $\sigma_8 \gtrsim 1$, is consistent with predicted
levels due to a secondary Sunyaev-Zeldovich anisotropy.  Further
observations are necessary to confirm the level of this excess and, if
confirmed, determine its origin.

\end{abstract}


\section{Introduction}
\label{sec:introduction}

Sub-horizon scale fluctuations in the Cosmic Microwave Background
(CMB) provide a direct view of simple, causal physical processes in
the early universe.  In standard cosmological models, the dominant
processes are acoustic oscillations of the primordial plasma and
photon-diffusive damping, which give rise to a harmonic series of
peaks in the CMB anisotropy spectrum modulated by an exponential
cutoff on small angular scales
\citep{silk,peebles_and_yu_1970,sunyaev_and_zeldovich_1970,bond_and_efstathiou_1987}.
Measurements of the CMB power spectrum in this regime provide strong
constraints on cosmological parameters \citep{White_et_al}, determine
the nature and initial conditions of the fluctuations
\citep{hu_and_white}, and provide fundamental tests of particle
physics \citep{KK}.  A number of experiments have detected peaks in
the anisotropy spectrum \citep{Miller_et_al_1999,boom,
maxima,new_maxima,dasispectrum,newboom}, which constrain the
cosmological models through measurements of the first, second, and
possibly the third acoustic peaks.  Observations at high multipoles
($\ell \sim 500 \rightarrow 2000$), where the physics is strongly
affected by photon diffusion and the thickness of the last scattering
region, provide independent constraints on these fundamental
parameters.  At even higher multipoles ($\ell > 2000$), secondary
effects such as the Sunyaev-Zeldovich Effect
\citep[SZE;][]{sunyaev_and_zeldovich_1972} are expected to dominate
\citep{rephaeli81,coleandkaiser} and hence offer the prospect of
studying the formation of large-scale structure at recent times.

This paper is one in a series reporting results from the Cosmic
Background Imager (CBI).  We have previously reported results between
$\ell = 600$ and $\ell = 1200$ in which we detected a damping tail in
the spectrum \citep[hereafter Paper I]{padin_cmb}.  These measurements
support the standard theoretical models of CMB anisotropies.  In the
present paper, Paper II, we present measurements with the CBI taken
from January through December of 2000.  These observations extend our
determination of the CMB anisotropy spectrum out to $\ell = 3500$.
The results are derived from deep integrations on three pairs of
pointings. We present a brief overview of the instrument and site in
\S~\ref{sec:instrument} and discuss our observing technique in
\S~\ref{sec:observations}. The data analysis methods are presented in
\S~\ref{sec:analysis}, including, in
\S~\ref{subsec:maximumlikelihood}, a discussion of the
maximum-likelihood analysis used to determine the power spectrum and,
in \S~\ref{subsec:sources}, a discussion of the method of removing the
discrete source foreground.  We present our results in
\S~\ref{sec:results}, and review our conclusions in
\S~\ref{sec:conclusion} .

The remaining papers in the series cover the CBI mosaic power spectrum
\citep[hereafter Paper III]{cbi_mosaic}; the implementation of the
maximum likelihood analysis \citep[hereafter Paper IV]{cbi_methods};
the cosmological interpretation of our results \citep[hereafter Paper
V]{cbi_cosmology}; and a possible interpretation of the excess power
observed at high-$\ell$ \citep[hereafter Paper VI]{cbi_excess}.

\section{The Instrument and Site}
\label{sec:instrument}

A detailed description of the CBI can be found in the paper by
\citet{padin_telescope}, so we summarize here only important aspects
of the instrument design.  The CBI is a planar synthesis array of 13
0.9-meter diameter Cassegrain antennas mounted on a 6-meter diameter
tracking platform.  Short baselines are susceptible to contamination
from cross-talk, so the antennas are surrounded by cylindrical shields
which provide more than 110 dB of isolation.  Further rejection of
cross-talk is accomplished through the use of differenced observations
(see \S \ref{sec:observations}).  A key feature of the CBI is its
sensitive broadband receivers, which are based on indium phosphide
high electron mobility transistor (HEMT) amplifiers operating in the
26--36 GHz band. The system temperatures measured on the telescope,
including CMB, ground, and atmosphere, are typically $\sim 30 \Kel$.
The receiver outputs are combined in an analog correlator with 10
1-GHz bands.  A $ 1 \, {\rm kHz}$ phase switching scheme is used to
reject cross-talk in the electronics.

The antenna platform is on a 3-axis mount with azimuth, elevation, and
parallactic angle axes; the elevation is restricted to $> 42^{\circ}$.
The parallactic angle rotation provides the ability to track the
rotation of the sky throughout the observation so that the $(u,v)$
orientation of a baseline on the sky is fixed, where $u$ and $v$ are
the orthogonal components of the baseline length measured in
wavelengths.  The platform is moved through additional discrete steps
in orientation to increase the $(u,v)$ coverage, test for false
signals, and increase ground rejection.  The antennas can be placed in
different locations on the platform, allowing the configuration to be
matched to a range of science goals.

The CBI is located at an altitude of 5080 m in the Atacama desert of
northern Chile, near the proposed site for the Atacama Large
Millimeter Array \citep{radford_site}.  This high dry site was chosen
in order to reduce atmospheric brightness fluctuations which would
otherwise limit the sensitivity of the CBI.  During the calendar year 2000,
$\sim 35\%$ of the nights were lost to weather and $\sim 5\%$ to
equipment malfunction.  Observing conditions were best from mid-August
through mid-December, when no nights were lost to weather.  In the
 nights when observing conditions were good, there
is no evidence of atmospheric effects in the CBI data except for rare
instances which typically correlate with the appearance of visible,
low-lying clouds.

\section{Observations}
\label{sec:observations}

The data for the results presented here were obtained from 2000
January 11 to December 12.  During this period both deep observations
of individual fields and mosaiced observations of multiple pointings
were made.  We present in this paper the results for three deep
fields, two of which are part of the mosaics; the CBI mosaic data,
images, and power spectra are discussed in Paper III.  We refer to
these three deep fields collectively as the CBI deep fields, and
individually as deep 08h, deep 14h, and deep 20h.  We obtained 42, 9,
and 49 good nights of observing on these fields.  Thus most of our
results come from the deep 08h and deep 20h fields, with the deep 14h
providing a somewhat weaker independent check.  All of the deep fields
were chosen to have low point-source contamination levels using the
NRAO Very Large Array Sky Survey \citep[NVSS]{condon}.

In order to eliminate the influence of solar radiation in distant
sidelobes, observations were made only at night.  For similar reasons,
no observations were used for which the field was less than
$60^{\circ}$ from the moon.  Our results are insensitive to the
precise lunar cut beyond $60^{\circ}$.  As discussed in Paper I, the
CBI has no ground shield and observations are differenced in order to
remove signals due to ground pickup on the short baselines.  To
accomplish this, two fields, designated {\it lead} and {\it trail},
separated by 8 minutes in right ascension, are tracked across almost
identical ranges of azimuth and elevation.  Ground pickup and other
contaminating signals are cancelled in the ${\it lead} - {\it trail}$
difference as long as they are stable over an 8 minute time-span.  The
potential residual of this in the differenced data has been shown in
Paper I to be $< 1.3\%$ of the cosmic signal.

\begin{deluxetable*}{l l l l l l}
\tabletypesize{\scriptsize}
\tablewidth{0pt}
\tablecaption{CBI Field Coordinates\label{tbl:field_coords}}
\tablehead{\colhead{Name}&
\colhead{${\rm R.A.}_{\rm {\it lead}}$ (J2000)}&
\colhead{${\rm R.A.}_{\rm {\it trail}}$ (J2000)}&
\colhead{Decl. (J2000)}&
\colhead{$b$}&
\colhead{$t_{\rm int}$}}
\startdata
deep 08h & 08:44:40  & 08:52:40  &   -03:10:00    & $24^{\circ}$ &$  131.2 \, {\rm h} $\\
deep 14h & 14:42:00  & 14:50:00  &   -03:50:00    & $40^{\circ}$ &$  24.0 \, {\rm h} $\\
deep 20h & 20:48:40  & 20:56:40  &   -03:30:00    & $-28^{\circ}$ &$  115.7 \, {\rm h} $\\
\enddata
\end{deluxetable*}

The coordinates of the {\it lead} and {\it trail} fields for the 3 CBI
deep fields, the total integration times on {\it lead} and {\it trail}
combined, and the Galactic latitudes are shown in
Table~\ref{tbl:field_coords}.  The integration times shown are
computed after all data filters have been applied.

\begin{figure}
\plotone{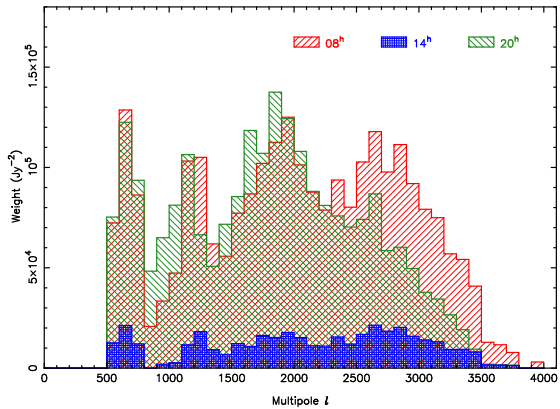}
\caption{The distributions of the weights of the visibilities as a
  function of $\ell$ for the three deep fields.  Since the CBI
  sensitivity does not vary much with baseline or frequency, these
  weights are roughly proportional to the on sky integration time as a
  function of $\ell$.}
\label{fig:wtplot}
\end{figure}

For the observations reported in this paper, four antenna
configurations were used.  The first configuration, on which the
results of Paper I were based, was a ring configuration, with most of
the receivers near the perimeter of the platform.  This provided
fairly uniform $(u,v)$ coverage, many long baselines for robust
calibration, and good access to the antennas and instrumentation on
the platform during the testing which accompanied initial
observations.  The second and third configurations were better
optimized for the CBI mosaic observations.  These more compact
configurations provided higher sensitivity in the $\ell = 600$--$1500$
range, and were well-suited to low-redshift Sunyaev-Zeldovich
observations.  In addition, the third configuration provided nearly
redundant baselines at different frequencies which enabled us to make
a more accurate determination of the radio frequency spectrum of the
signals reported in Paper I.  The last configuration was an extended
configuration with more uniform coverage than the first, more long ($>
5 \, {\rm m}$) baselines, which are useful for point source
monitoring, and greater sensitivity for the frequency spectrum
determination out to $\ell = 1400$.  The distributions of the weights
of the data as a function of $\ell$ is shown in
Figure~\ref{fig:wtplot}.  The 08h field was observed in the first,
second and fourth configurations; the 14h field in the first
configuration; and the 20h field in the third configuration.

In the first three configurations, all but one of the receivers
were configured to be sensitive to left circular polarization (LCP);
the remaining receiver was sensitive to right circular polarization
(RCP).  In the last configuration all antennas were configured for LCP.

\section{Data Analysis}
\label{sec:analysis}

\subsection{Calibration and Editing}
\label{subsec:calib_edit}

The equatorial sky coverage and relatively large antennas of the CBI
enable us to use a variety of celestial objects as calibrators,
including planets, supernova remnants, radio galaxies and quasars.
This is convenient for comparing the CBI flux density scale with other
measurements (see below) and for accurately calibrating the data.  CBI
observations are calibrated using nightly measurements of primary
calibrators, chosen for their strength and lack of variability at
these frequencies (Jupiter, Saturn, Tau A [3C144, the Crab Nebula],
and Vir A [3C274]), and a set of secondary calibrators (3C279, 3C273,
J1743$-$038, B1830$-$210, and J1924$-$292).  Our estimates of the
primary calibrator flux densities are based on CBI measurements of
these sources relative to Jupiter assuming an apparent Rayleigh-Jeans
brightness temperature $T_{\rm Jup} = 152 \pm 5 \, {\rm K}$ at 32 GHz
\citep{Mason_et_al_1999}. This value is uncorrected for the
occultation of the CMB by the planet, and applying this correction
yields a Rayleigh-Jeans brightness temperature of $154 \pm 5 \, {\rm
K}$.  This is also the basis of the Very Small Array calibration
\citep{vsa_calib}.  Since Jupiter is known to have a non-thermal
spectrum at these frequencies, and the precise spectral shape is not
independently well-determined, this flux density scale is transferred
to each of the CBI bands by observations of Tau A, for which the
\citet{baars_1977} spectral index $\alpha=-0.299$ is assumed, where $S
\propto \nu^{\alpha}$.  There is good evidence that the power law
index extends to high frequencies \citep[e.g.,][]{mezger1986}.  The
flux densities of the secondary calibrators are bootstrapped from the
nearest primary calibrator observations, and are used when the primary
calibrators are either not visible or are too close to the moon.
Observations of the primary calibrators over the year 2000 indicate
random calibration errors of $\sim 3\%$ night-to-night. The residuals
of a polynomial fit to the light curves of the secondary calibrators
show a similar scatter.  In light of this and the $3.3\%$ calibration
uncertainty of the \citet{Mason_et_al_1999} value for the temperature
of Jupiter we assign a calibration uncertainty of $5\%$ to our data.
This corresponds to a $10\%$ uncertainty in bandpower (${\rm \mu
K^2}$).  These include the beam uncertainties discussed below.

We have two comparisons of our flux density scale with that of the NRAO
Very Large Array\footnote{The National Radio Astronomy Observatory is
a facility of the National Science Foundation operated under
cooperative agreement by Associated Universities, Inc.} (VLA) at 22.4
GHz.  First, we compare CBI observations of Mars with the model for
the Mars temperature which is used at the VLA \citep{rudy_1987},
improved by M. Gurwell and B. Butler (private communication, February
2002).  R. Perley (private communication, February 2002) has shown
that the flux density scale at 22.4 GHz based on this Mars temperature
model agrees to within $0.4\%$ with the flux density scale determined
at this frequency based on a model of NGC7027 derived by A. van Hoof
(private communication, February 2002).  The ratio of the temperature
of Mars from the CBI to that of the improved Rudy model is $ 1.026\pm
0.013$. Our second comparison is based on observations of 3C273 on the
CBI and the VLA at 7 epochs for which the time interval between
observations on the two instruments was small enough to eliminate
uncertainties due to source variability.  The VLA observations have
been calibrated using the improved Rudy Mars model at 22.4 GHz, and
the CBI 26--36 GHz measurements have been extrapolated to this
frequency.  The ratio of the flux density of 3C273 on the CBI to that
on the VLA is $0.984 \pm 0.010$. From the weighted averages of these
two tests we find that the CBI flux density scale, based on the
\citet{Mason_et_al_1999} absolute calibration, is $0.1\% \pm 0.8\%$
lower than the VLA flux density scale at 22.4 GHz.  The level of
agreement between these two scales is clearly fortuitous, but it does
give us confidence that the $5\%$ calibration uncertainty we estimate
for the CBI flux density scale is, if anything, conservative.

Individual scans of calibrators and the microwave background are
bracketed by measurements of an internal noise calibration source
whose equivalent flux density for each baseline and channel is
determined by reference to the celestial flux density calibrators. The
noise calibration source was initially intended to remove instrumental
gain fluctuations over the course of a night, but we found that the
instrumental gains were more stable ($\sim 3\%$ rms variations) than
the noise calibration source output and so this was not used.  This is
primarily because of the difficulty in stabilizing the temperature of
the noise calibration source amplifier and various other components,
all of which lie outside of the cryogenic dewars.  All of the
(complex) noise calibration source measurements for a given night are
averaged together and the data from all baselines are scaled to give
an identical response.  This removes baseline-based gain and phase
calibration errors; it also introduces antenna-based errors which are
removed by the subsequent (antenna-based) primary flux density
calibration.  At the start and end of each night of observing the
relative gain and phase errors between the real and imaginary channels
were measured using the noise calibration source with and without a
$90^{\circ}$ phase shift applied to the receiver local oscillator.
The rms quadrature phase is $\sim 5^{\circ}$ and the rms gain errors
are $\sim 10\%$.  The solutions are stable over a timescale of several
weeks.

{\it Lead} and {\it trail} scans of the CMB fields are interleaved
with 1 minute observations of bright ($> 1 \, \Jy$) nearby calibrators
which provide a check on the telescope pointing.  From these
observations we determine that the absolute rms radio pointing is $\sim 22''$,
while the rms tracking errors are $\sim 2''$.

Data are flagged by both automatic and manual filters. The on-line
control system identifies corrupt or potentially unreliable data.
These affect periods when the telescope was not tracking properly, a
receiver was warm, a local oscillator was not phase locked, the total
power of a receiver was out of the normal range, or when a receiver
phase-shifter was not acquired.  Observer notes were also used to
examine periods when there were instrumental problems that might have
affected the data quality or there was bad weather, identified by
visible cloud cover or corrupted visibilites on the short baselines.
A small fraction ($\sim 2\%$) of the data were deleted manually on the
basis of these inspections.  Subsequent automatic data edits are based
on the data scatter and baseline-to-baseline correlations.  The first
level of rejection is a $5\sigma$ scan-wise outlier edit to eliminate
occasional instrumental glitches which give rise to large signals in
individual samples.  This cut affects a negligible fraction of the
data, and our results are not sensitive to the precise level of this
cut.

Further automated data-filtering is provided by our differencing
procedure, which also generates an accurate estimate of the thermal
noise in each scan for each baseline and frequency channel.  To do the
differencing, individual 8.4-s integrations in the {\it lead} and {\it
trail} scans, which are taken 8 minutes apart, are matched and
subtracted.  Since our fields are separated by 8~min in observing
time, the typical scan length is $\sim 400$ s after slewing and firing
the noise calibration source, corresponding to $\sim 50$ integrations
of 8.4 s each. In order to prevent a few short scans from biasing the
statistics, any scan with fewer than 30 pairs of 8.4 s integrations
was rejected.  The noise estimates are derived from the differences
between the $lead$ and $trail$ fields for individual scans of $\sim
50$ 8.4 sec integrations.  Data from different baselines and times
which contribute to the same $(u,v)$ points are subsequently combined.
This leads to an underestimate of $\sim 8$\% in the variance (see
Appendix) which we correct for in our statistical analyses and
spectrum determinations.

The celestial signals are stationary on an 8-minute timescale, and
constant ground signal is cancelled by the ${\it lead}-{\it trail}$
differencing.  The noise will be increased by atmospheric noise,
ground signal variations, or previously unflagged instrumental
anomalies. On this basis any scan for which the rms of the samples is
greater than twice the expected thermal noise level is rejected: this
primarily eliminates periods affected by weather.  Less than $0.1\%$
of the data are rejected by this filter.  The noise level is
established with reference to data which are clearly free of
atmospheric contamination; it is highly repeatable and within $\sim
20\%$ of the expected thermal noise level.

In computing the data differences and thermal noise level, we also
compute the covariance between all baseline-channel pairs, i.e.,
the thermal noise covariance matrix.  For the deep 08h field, a single
night (2000 January 12) was found to have statistically significant
off-diagonal correlations.  On this night there were clear instances
of atmosphere emission on the short baselines, and the baseline
correlations are likely due to poor weather.  This day was excluded
from the analysis.  No anomalous correlations are seen in the deep 14h
data.  The correlation test on the deep 20h data revealed a single
baseline-channel with a large real-imaginary anti-correlation.  When
this baseline-channel was eliminated from the 20h dataset, no
significant baseline-baseline correlations are seen.  The excision of
these data from the 08h and 20h fields has a negligible effect on the
measured power spectrum, suggesting that lower-level, undetected
correlator problems of a similar nature also do not affect our result.

The telescope primary beam has been measured accurately using
observations of Tau A.  The primary beam width (FWHM) is $45.2\times
(31\;{\rm GHz}/\nu)$ arc minutes; we estimate a $0.7\%$ uncertainty in
the beamwidth.  This gives rise to a $1.4\%$ contribution to our
overall calibration error.  Note that unlike the case for total power
experiments, uncertainties in the primary beam width for an
interferometer affect only the overall temperature scale, and do not
introduce an $\ell$-dependent systematic error.  The beam-width
measurement is described in Paper III.

\subsection{Maximum Likelihood Analysis}
\label{subsec:maximumlikelihood}

\subsubsection{Formalism and Algorithms}

The determination of the angular power spectrum of the sky from
interferometer measurements has been discussed by \citet{hobson_et_al}
and \citet{white_et_al_1999}; the more general framework of power
spectrum estimation is discussed by \citet{bond_jaffe_knox} and
\citet{bjk2}. Here we present a brief overview of our procedure. The
details of the method are given in Paper IV. We test a space of
hypothetical models for the spectrum, $C_{\ell}$, against the data
using the likelihood function as a figure of merit.  For complex data
points $\vect{V}$ with zero mean and Gaussian noise described by the
covariance matrix ${\bf C}_{jk}$, this is
\begin{equation}
 L(C_{\ell})= {1 \over {\pi^n \, {\rm det}\,{\bf C}}}\; {\rm exp}\,\{{\vect{-V^*}}
 {\bf C}^{-1}{\vect {V}}\}.
\end{equation}
where the covariance matrix, $\tjptensor{C}$, is given by 
\begin{equation}
\tjptensor{C} = \tjptensor{C}^{\rm N} + \sum_B q_B \tjptensor{C}_B^{\rm S} 
                          + q_{\rm src} \tjptensor{C}^{\rm src}
                          + q_{\rm res} \tjptensor{C}^{\rm res}.
\label{eq:cvm}
\end{equation}
Here $\tjptensor{C}^{\rm N}$ is the noise covariance and
$\tjptensor{C}_B^{\rm S}$ is the sky (or theoretical CMB) covariance for
band $B$.  The foreground covariance is made up of $\tjptensor{C}^{\rm
src}$, a constraint matrix for point sources with known positions, and
$\tjptensor{C}^{\rm res}$, a matrix which models the contribution of
faint point sources of unknown position.  The calculation of the
expected sky variance $\tjptensor{C}_B^{\rm S}$ is discussed in many
references (e.g., \citealt{hobson_et_al}; Paper I; Paper IV).  For a
well-designed interferometer the noise term $\tjptensor{C}^{\rm N} $ is
diagonal (see also \S~\ref{subsec:calib_edit}).  The pre-factors
$q_{\rm res}$ and $q_{\rm src}$ are held fixed at {\it a priori}
values (see \S~\ref{subsubsec:statisticalsources}).  As discussed
below, the $q_B$ are varied to yield estimates of the microwave
background bandpower.

\begin{figure}
\plotone{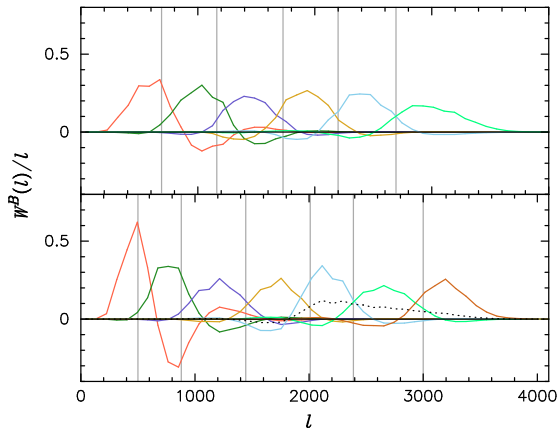}
\caption{CBI joint Deep Field window functions.  The alternate binning
window functions are displayed in the upper panel; the primary binning
window functions are displayed in the lower panel.  Vertical lines
denote the bandpower bin divisions discussed in the text.  The window
function of the high-$\ell$ rebinning is shown as a dashed line. The
expected value of band power for a given spectrum, $C_l$, in one bin
is $\Sigma_l\; W(l)/l \times l(l+1)C_l/2\pi$ }
\label{fig:deepwins}
\end{figure}

For the present analysis of the deep field data we adopt a parametric
description of the $C_{\ell}$ as seven bins with lower boundaries at
$\ell = 0$, 500, 880, 1445, 2010, 2388, and 3000.  Within each bin,
the power spectrum is assumed to be flat in $\ell (\ell +1) C_{\ell}$.
The overall amplitudes of the bins $q_B$ are treated as free
parameters in the maximization of the likelihood.  The spacing of
$\Delta \ell$ is set by the intrinsic resolution of the CBI in the
aperture plane (${\rm \Delta{\ell}} \sim 300$).  Here $\Delta \ell$
is the FWHM of the visibility window function (see Paper III).  This
resolution can be increased by mosaicing.  The divisions above were
chosen to coincide with breaks in the multi-configuration $(u,v)$
coverage, which minimize correlations between adjacent bins.  Given
the $(u,v)$-coverage of our data, this choice of binning yields $10$--$15\%$
anti-correlations between adjacent bins. To demonstrate that our
results are not sensitive to the choice of bins, we have also
conducted the analysis with an alternate set of bins whose boundaries
are halfway between those of the primary binning described above (six
bins with lower boundaries at $\ell = 0$, 690, 1162, 1728, 2199, and
2694).  Window functions for each band are shown in
Figure~\ref{fig:deepwins}. These show the sensitivity of a given
bandpower bin to power in individual multipoles as a function of
$\ell$.

We compute the parameter uncertainties and correlations from the
curvature of the likelihood at the best-fit locus.  In practice we use
the Fisher matrix
\begin{equation}
 F_{BB'} = \frac{1}{2} {\rm Tr} \left( {\bf C}^{-1} \frac{\partial {
 \tjptensor{C}^{\rm S}}}{\partial q_{B}}  {\bf C}^{-1} \frac{\partial {
 \tjptensor{C}^{\rm S}}}{\partial q_{B'}}\right)
\end{equation}
as a computationally efficient estimator of the curvature.  In the
approximation that the likelihood function is Gaussian, $F_{BB'}^{-1}$
is the correlation matrix of the fitted parameters.  As a check on the
Fisher matrix results, we have also mapped the likelihoods in
individual bands and checked against the error bars computed in the
offset log-normal approximation \citep{bjk2}.  We find good agreement
between these three methods.  Monte Carlo tests of simulated CBI data
show that these error estimates, as well as the bandpower estimates
themselves, are unbiased.

The dominant foreground for the CBI is composed of discrete
extragalactic radio sources described by the correlation matrices
$\tjptensor{C}^{\rm res}$ and $\tjptensor{C}^{\rm src}$.  For purposes of
this analysis we neglect other possible foregrounds, although see
\S~\ref{sec:results} for limits on the possible contribution of
foregrounds on large angular scales.

\subsubsection{Implementation}

Since in the small-angle limit interferometers directly measure linear
combinations of the sky Fourier modes, the extraction of angular power
spectra from the data is in principle simpler than in the case of
total power experiments.  In general, however, $\tjptensor{C}_{jk}$ has
non-negligible off-diagonal elements, and depending on the
$(u,v)$-sampling of the dataset this matrix can be large.  To deal with
the computational challenge of inverting this matrix, we have
developed a Fourier gridding algorithm.  This algorithm constructs a
data vector composed of linear combinations of the visibility data and
calculates the resulting covariance matrix. For single-field analyses
such as we report here, the primary effect of this gridding is to
reduce redundancies in the $(u,v)$-dataset caused by dense aperture
plane sampling. This compression is typically a factor of 3 to 4: for
each field there are several thousand visibilities, which are gridded
down to 970 estimators.  This results in over an order of magnitude
increase in the speed of the likelihood analysis.  This pipeline and
the associated formalism are discussed in Paper IV.

The pipeline was implemented on the 4 and 32 processor shared memory 
Alpha GS320 and ES45 clusters at CITA. The codes were parallelized using Open
MP directives. A complete joint analysis of the three deep fields
takes approximately 4 CPU hours.  This includes pre-gridding and the
calculation of all ancillary quantities such as the bandpower window
functions, likelihood maps and noise spectra. Bandpower estimates can
be obtained in as little as 1 CPU hour for the deep field data.  These
rapid turnaround times permitted extensive testing of the data and
pipeline; results of some of these tests are discussed in \S
\ref{subsec:consistency}.

\subsection{Discrete Sources}
\label{subsec:sources}

Extragalactic radio sources are the dominant foreground to CMB
observations over the 26--36 GHz frequency band on arcminute scales
\citep{Tegmark_and_Efstathiou}.  If no allowance for this foreground
is made, the few brightest discrete sources in each field dominate any
power spectrum determination beyond $\ell \sim 1000$.  While much is
known about radio source populations at lower frequencies, this is not
the case at 31 GHz, and observations at this frequency are essential
in dealing with the point source contamination at high $\ell$.  We
therefore constructed a 26--34 GHz receiver for the OVRO 40 meter
telescope which we have used to study the source population at this
frequency and to provide a direct check on the CBI calibration.  We
have also computed 31 GHz source counts directly from the CBI deep
maps, and the mosaic maps presented in Paper III.  The results from
these complementary approaches are presented in
\S~\ref{subsubsec:brightsources}.  These results form the basis of our
strategy for dealing with both the bright sources and the statistical
residual source background.  The treatment of radio sources in the CMB
analysis is detailed in \S~\ref{subsubsec:statisticalsources}.

\subsubsection{31 GHz Radio Source Measurements}
\label{subsubsec:brightsources}

\begin{deluxetable*}{lllllll}
\tabletypesize{\scriptsize}
\tablewidth{0pt}
\tablecolumns{7}
\tablecaption{Subtracted Discrete Sources\label{tbl:sourcetable}}
\tablehead{\colhead{Name}&
\colhead{R.A. (J2000)}&
\colhead{Decl. (J2000)}&
\colhead{\begin{tabular}{c}$S_{31}$\\ (mJy)\end{tabular}}&
\colhead{Field}&
\colhead{\begin{tabular}{c}$r$\\ (arcmin)\end{tabular}}& 
\colhead{\begin{tabular}{c}$S_{\rm CBI}$\\ (mJy)\end{tabular}}}
\startdata
 084204-0317 & 08:42:04.110 & -03:17:06.9  &  $6.0 \pm 1.4$   &C08(L) & 39.7 & 0.5 \\
 084242-0344 & 08:42:42.690 & -03:44:25.8  &  $8.7\pm  1.9$  &C08(L) & 45.3 & 0.3 \\
 084336-0302 & 08:43:36.940 & -03:02:59.5  &  $7.9\pm  1.2$ &C08(L) &  17.3 & 5.3 \\
 084533-0217 & 08:45:33.200 & -02:17:31.4  &  $6.4\pm  1.5$  &C08(L) &  54.1 & 0.01 \\
 084553-0305 & 08:45:53.200 & -03:05:38.4  &  $9.1\pm  1.8$  &C08(L) &  18.8 & 5.7 \\
 084553-0342 & 08:45:53.740 & -03:42:02.3  &  $9.0\pm  2.0$  &C08(L) &  37.0 & 1.2 \\
 084730-0251 & 08:47:30.490 & -02:51:36.6  &  $6.5\pm  1.2$ &C08(L) &  46.5 & 0.2 \\
 084732-0340 & 08:47:32.920 & -03:40:39.7  &  $42.6 \pm 1.9$ &C08(L) &  53.0 & 0.1 \\
 084803-0257 & 08:48:03.110 & -02:57:52.4  &  $20.7 \pm 1.6$  &C08(L) &  52.3 & 0.1 \\
 084944-0317 & 08:49:44.490 & -03:17:59.2  &  $47.3 \pm 1.9$ &  C08(T) &  44.7 & 1.8 \\
 085311-0342 & 08:53:11.900 & -03:42:48.1  &  $8.0 \pm 1.4 $ & C08(T)&  33.8 & 1.6 \\
 085322-0259 & 08:53:22.310 & -02:59:48.7  &  $7.8 \pm 1.4 $ & C08(T)&  14.7 & 5.9 \\
 085326-0211 & 08:53:26.340 & -02:11:49.3  &  $9.4 \pm 2.3 $ & C08(T)&  59.3 & 0.01 \\
 085328-0341 & 08:53:28.250 & -03:41:08.0  &  $41.4\pm  1.1$ & C08(T)&  33.4 & 8.4 \\
 085329-0258 & 08:53:29.860 & -02:58:04.3  &  $7.0 \pm 1.6$  &C08(T) &  17.3 & 4.7 \\
 085359-0302 & 08:53:59.440 & -03:02:56.6  &  $15.0 \pm 2.6$ &C08(T) &  21.1 & 8.3 \\
 085430-0223 & 08:54:30.480 & -02:23:14.9  &  $17.8 \pm 1.5$   &C08(T) &  54.3 & 0.03 \\
 144019-0308 &  14:40:19.430 &  -03:08:34.5 &   $8.5 \pm   1.7$ & C14(L) &  48.5 & 0.1 \\
 144042-0406 &  14:40:42.950 &  -04:06:47.3 &   $9.4 \pm   1.7$  & C14(L) &  25.6 & 3.9 \\
 144254-0329 &  14:42:54.290 &  -03:29:34.8 &   $7.0 \pm   1.4$  & C14(L) &  24.5 & 3.1 \\
 144400-0259 &  14:44:00.070 &  -02:59:20.3 &   $6.4 \pm   1.5$  & C14(L) &  58.9 & 0.01 \\
 144444-0400 &  14:44:44.990 &  -04:00:47.7 &   $9.8 \pm   1.8$  & C14(L) &  42.7 & 0.5 \\
 144445-0311 &  14:44:45.870 &  -03:11:51.1 &   $9.1 \pm   1.8$  & C14(L) &  56.4 & 0.01 \\
 144457-0312 &  14:44:57.220 &  -03:12:03.4 &   $30.9 \pm   3.0$ & C14(L) &  58.4 & 0.03 \\
 144459-0311 &  14:44:59.650 &  -03:11:04.1 &   $11.9 \pm   2.4$  & C14(L) &  59.5 & 0.02 \\
 144506-0326 &  14:45:06.230 &  -03:26:13.4 &   $15.5 \pm   3.3$  & C14(L) &  52.4 & 0.06 \\
 144542-0329 &  14:45:42.430 &  -03:29:57.4 &   $41.3 \pm   2.9$  & C14(L) &  59.2 & 0.05 \\
 144824-0316 &  14:48:24.510 &  -03:16:47.1 &   $6.1 \pm   1.4$  & C14(T) &  40.9 & 0.5 \\
 144933-0352 &  14:49:33.660 &  -03:52:19.2 &   $22.1 \pm   3.7$ & C14(T) &  7.0 & 20.7 \\
 144954-0302 &  14:49:54.240 &  -03:02:36.7 &   $21.9 \pm   2.8$  & C14(T) &  47.4 & 0.4 \\
 145146-0356 &  14:51:46.180 &  -03:56:53.7 &   $26.1 \pm   1.8$  & C14(T) &  27.5 & 9.2 \\
 145207-0325 &  14:52:07.080 &  -03:25:04.1 &   $7.8 \pm   1.4$  & C14(T) &  40.4 & 0.6 \\
 145230-0315 &  14:52:30.250 &  -03:15:35.2 &   $6.7 \pm   1.0$  & C14(T) &  51.0 & 0.05 \\
 204542-0323 &  20:45:42.330 &  -03:23:03.6 &   $13.0 \pm   2.2$  & C20(L) &  45.0 & 0.5 \\
 204603-0336 &  20:46:03.810 &  -03:36:47.3 &   $11.1 \pm   1.8$  & C20(L) &  39.7 & 1.0 \\
 204608-0249 &  20:46:08.670 &  -02:49:39.2 &   $11.7 \pm   1.4$  & C20(L) &  55.4 & 0.01 \\
 204645-0316 &  20:46:45.400 &  -03:16:15.1 &   $9.2 \pm   2.0$ & C20(L) &  31.8 & 2.2 \\
 204710-0236 &  20:47:10.320 &  -02:36:22.7 &   $197.7 \pm   1.9$ & C20(L) &  58.1 & 0.2 \\
 204731-0356 &  20:47:31.600 &  -03:56:11.5 &   $6.4 \pm   1.4$  & C20(L) & 31.3  & 1.6 \\
 204745-0246 &  20:47:45.660 &  -02:46:05.0 &   $79.7 \pm   1.6$ & C20(L) &  46.0 & 2.3 \\
 204800-0243 &  20:48:00.130 &  -02:43:03.7 &   $9.4 \pm   1.1$  & C20(L) &  48.0 & 0.2 \\
 204830-0428 &  20:48:30.610 &  -04:28:20.3 &   $7.0 \pm   1.7$  & C20(L) &  58.4 & 0.01 \\
 205001-0249 &  20:50:01.360 &  -02:49:06.0 &   $14.9 \pm   2.4$ & C20(L) &  45.7 & 0.5 \\
 205038-0305 &  20:50:38.860 &  -03:05:59.1 &   $39.4 \pm   2.5$ & C20(L) &  38.2 & 4.5 \\
 205041-0249 &  20:50:41.330 &  -02:49:17.0 &   $15.9 \pm   3.3$ & C20(L) &  50.8 & 0.1 \\
 205045-0337 &  20:50:45.050 &  -03:37:43.3 &   $15.2 \pm   2.7$ & C20(L) &  32.3 &3.5 \\
 205355-0259 &  20:53:55.400 &  -02:59:43.2 &   $10.3 \pm   1.3$ & C20(T) &  51.2 & 0.06 \\
 205520-0306 &  20:55:20.170 &  -03:06:17.5 &   $6.0 \pm   1.4$ & C20(T) &  31.0 & 1.6 \\
 205543-0350 &  20:55:43.650 &  -03:50:51.6 &   $32.3 \pm   2.1$ & C20(T) &  25.2 & 13.6 \\
 205550-0416 &  20:55:50.260 &  -04:16:46.8 &   $180.7 \pm   4.5$ & C20(T) &  48.4 & 2.8 \\
 205735-0250 &  20:57:35.150 &  -02:50:49.5 &   $14.6 \pm   3.1$  & C20(T) &  41.5 & 1.0 \\
 205812-0312 &  20:58:12.410 &  -03:12:26.0 &   $6.3 \pm   1.0$ & C20(T) &  29.1 & 2.0 \\
 205822-0303 &  20:58:22.070 &  -03:03:26.8 &   $6.0 \pm   1.2$ & C20(T) &  36.9 & 0.8 \\
 210000-0343 &  21:00:00.370 &  -03:43:02.8 &   $11.0 \pm   2.4$ & C20(T) &  51.8 & 0.05 \\
 210007-0325 &  21:00:07.950 &  -03:25:43.7 &   $12.9 \pm   2.9$ & C20(T) &  52.3 & 0.05 \\
 210029-0317 &  21:00:29.510 &  -03:17:41.6 &   $8.8 \pm   2.1$ & C20(T) &  58.8 & 0.01 \\
\enddata
\end{deluxetable*}

The OVRO 40-meter telescope was used at 31 GHz to survey the $2225$
NVSS sources brighter than $S_{1.4 \, {\rm GHz}} = 6 \mJy$ in four
22.5 square degree fields which encompass the deep and mosaic
fields. With the $4\sigma$ cutoff specified below, the survey has a
$7\%$ chance of one or more false detections. To minimize the effects
of variations in the flux density of the monitored sources, the
40-meter and CBI observations were made as nearly as possible over the
same period of time. The typical sensitivity achieved on the 40-meter
telescope was $ 2 \, {\rm mJy} \, ({\rm rms})$, although the
distribution extends to $5 \, {\rm mJy}$. $12 \, \%$ of the targeted
NVSS sources were detected above the survey thresholds of $4\sigma$
and $S_{\rm min,31}=6 \, \mJy$.  The survey is $90\%$ complete at $S_{31}
> 16 \mJy$ and $99\%$ complete at $S_{31} > 21 \mJy$.  The 56 detected
sources within 1 degree of the {\it lead} or {\it trail} field centers
were subtracted directly from the CBI data; these sources are shown in
Table~\ref{tbl:sourcetable}. This table lists total flux densities,
$S_{31}$; distances in arc minutes from the center of the CBI field,
$r$; and the flux density predicted on the CBI including the effects
of the CBI primary beam, $S_{\rm CBI}$.  Few of these 56 sources
contribute significantly to the data since most of them are too
distant from the CBI field centers to do so: in each deep field only a
few sources are visible in the images.  The spectral indices of these
sources were computed by comparison with the NVSS 1.4 GHz flux
densities, and we find $\bar{\alpha}=-0.45$ with an rms dispersion of
0.37; the minimum and maximum spectral indices were $-1.32$ and
$+0.50$.

We have used the CBI deep and mosaic maps to determine the
source counts at 31 GHz.  To do this, we made maps from the long
baseline ($> 250 \, \lambda$) data and searched for peaks over the
$5 \sigma$ threshold.  Below this level the probability of false
detections is significant.  A fit to the resulting counts over the
range $5$--$50$ mJy yields
\begin{equation}
N(>S_{31}) = 2.8 \pm 0.7 \, {\rm deg^{-2}} \, \left(\frac{S_{31}}{10
  \, \mJy}\right)^{-1.0}
\end{equation}
This is slightly higher than the 31 GHz counts estimated from either
the NVSS counts or the 15 GHz counts of \citet{taylorcounts}.  In both
the deep and mosaic maps no sources were detected above the $5\sigma$
limit which were not NVSS objects.  The area searched depends on the
limiting flux density chosen: in the deep maps the area was 6 square
degrees at 12 mJy, falling to 1.8 square degrees at 6 mJy. In the
mosaic maps the area was 47 square degrees at 25 mJy, falling to 12.2
square degrees at 18 mJy.  The flux densities of the sources detected
in both the deep and mosaic maps are consistent with the OVRO
determinations, with the agreement typically within $5\%$.

As discussed above, the OVRO-detected bright sources were directly
subtracted from the CBI data since the OVRO source subtraction is a
step in the standard data-analysis pipeline which has been useful in
producing diagnostic images of the data.  In
\S~\ref{subsubsec:statisticalsources} we discuss a method (the
constraint matrix approach of \citealt{dasispectrum}) of dealing with point source foregrounds
which does not require knowledge of the source flux densities.  We
find that this method applied to the OVRO sources gives very similar
results as direct subtraction.  These complementary techniques of
dealing with this bright source population have been useful in
extending our power spectrum measurments to the instrumental limit of
3500.

\subsubsection{Statistical Treatment of Sources}
\label{subsubsec:statisticalsources}

In addition to the brightest sources in the CBI fields, which have
been measured and subtracted with the OVRO data, we must deal with
sources too faint to measure directly but which contribute
statistically to our measurements.  Since most radio sources have
spectra which fall towards higher frequencies, and since there are
wide-area radio sky surveys at lower frequencies with sensitivities
comparable to what we have in the CBI deep field maps, most of the
sources which contribute are readily identified from the low-frequency
surveys.  For this purpose we use the NVSS. There is also a small
contribution due to sources which are too faint to appear even in
NVSS.  We call these latter ``residual'' sources and estimate their
contribution from the CBI 31 GHz source counts.

After the subtraction of the bright sources monitored at OVRO, there
remain three components of point source contamination: one due to data
correlations induced by the (imperfect) OVRO source subtraction; a
component due to the NVSS sources whose positions are known, but whose
flux densities are not measured at 31 GHz; and a residual component
due to sources whose contribution must be estimated from source
counts.  These are represented as
\begin{equation}
\label{eq:srcmarg}
q_{\rm OVRO} \tjptensor{C}^{\rm OVRO}+ q_{\rm NVSS} \tjptensor{C}^{\rm NVSS} +
q_{\rm res} \tjptensor{C}^{\rm res}
\end{equation}
The $q$ are prefactors which are held fixed in the bandpower fit, and
allow for the effect of overall uncertainties in our estimates of the
source covariances. If $q=1$, an extra amount of variance equal to the
variance assumed in the calculations of the source covariance matrices
is allowed in each source mode. In the limit $q \rightarrow \infty$
the data modes corresponding to the point sources are completely
removed from the data, {\it i.e.}, these modes are ``projected out''
\citep{dasispectrum,bond_jaffe_knox}.  We note that it is possible to
allow the $q$ to be free parameters in the fit, and we do not find
significantly different results in this case; however it is more
conservative to fix the $q$ at very large values.  Explicit procedures
for calculating the three source covariances matrices are discussed in
Paper IV.

We adopt the strategy of projecting the modes corresponding to OVRO
residuals and NVSS sources out of our data.  A nominal variance is
assigned to each source based on the OVRO measurement uncertainty or,
for the NVSS sources, the width of the OVRO-determined 1.4 to 31 GHz
spectral index distribution.  We then set $q_{\rm OVRO} = q_{\rm NVSS} =
30$ which projects these modes out of the data.  We find that our
power spectra at high $\ell$, where point sources are most
troublesome, are little affected by variations in $q$ between $q=30$
to $q=10^6$.

At significantly larger values of $q$ the matrices are
numerically ill-conditioned.  We have tested the source projection
algorithm extensively on both real and simulated data and find that it
is robust against, for instance, randomly reassigning estimated
variances between sources and randomly perturbing the source positions
assumed in the projection by $30''$ (rms).

We model the contribution of the residual sources ($\tjptensor{C}^{\rm
res}$) as a Gaussian white-noise foreground. We determine the
normalization of this matrix from Monte Carlo simulations assuming the
source counts presented in \S~\ref{subsubsec:brightsources} to
generate random populations at 31 GHz. Since at fainter flux levels
the counts are expected to flatten somewhat, we used a slightly
shallower slope for the counts ($N(>S) \propto S^{-0.875}$) for this
calculation, although at the $1\sigma$ level our result is consistent
with that obtained for $N(>S) \propto S^{-1}$.  In order to determine
what fraction of these sources fall below the NVSS detection threshold
of 3.4 mJy and hence are residual, the sources are extrapolated down
to 1.4 GHz using the observed OVRO-NVSS distribution of spectral
indices (with $\bar{\alpha}=-0.45$ --- see
\S~\ref{subsubsec:brightsources}). Since the OVRO survey is strongly
biased towards flat-spectrum sources this will overestimate the
residual source power level. We find a 31 GHz variance of $C_0^{\rm
res} = 0.08 \, {\rm Jy^2\,sr^{-1}}$.  We get a similar result,
$C_0^{\rm res} =0.10 \, {\rm Jy^2\, sr^{-1}}$ from the analytic
calculation described in Paper IV.  In our highest bin this
corresponds to a power level of $\sim 115 \, {\rm \mu K^2}$, which is
less than the thermal noise in this bin and all others; we assign an
uncertainty of $50\%$ to this value, corresponding to $\pm 0.04\, {\rm
Jy^2\,sr^{-1}}$ ($57 \, {\rm \mu K^2}$).  This uncertainty is based on
varying the number counts and spectral index parameters within the
range allowed by their uncertainties, and rerunning the Monte Carlo
simulations used to calculate the residual power. Comparable power
levels are obtained if, instead of using the 31 GHz counts, we
extrapolate in frequency from the 1.4 GHz number counts using a simple
Gaussian model for the spectral index distribution which is
constrained to reproduce the detection statistics observed in the OVRO
survey.  As a test on the robustness of this correction we re-ran the
power spectrum analysis doubling the NVSS/residual source threshold
from 3.4 to 6.8 mJy at 1.4 GHz with similar results.  The number
density of the residual sources on the sky is too high to permit full
projection of these data modes, and in any case their positions are
not known so this would be impossible. We therefore fix $q_{\rm
res}=1$, which approximately corresponds to subtracting our
conservative estimate of the residual source power level from the
spectrum.

The cost of projecting out the NVSS sources is a loss of sensitivity
of the CBI at high $\ell$ due to our limited knowledge of the faint
point sources at 31 GHz.  The best way to address this problem in the
near term is with a sensitive 31 GHz receiver on the NRAO Green Bank
Telescope (GBT).  Such an instrument is under construction and should be in
use by the end of 2003.

\subsection{Data Consistency Tests}
\label{subsec:consistency}

The redundancy of observations on the CBI deep fields allows many
checks on the dataset.  Below we describe the key tests performed.

\subsubsection{$\chi^2$ Tests}
\label{subsubsec:chi2}

A simple and direct test of the data consistency is provided by
computing the $\chi^2$ of visibilities with the same $(u,v)$
coordinate, frequency, and sky pointing.  We compared the data at each
$(u,v)$ point on each night with the average of all other nights.
This procedure identified three days on the deep 20h field which were
formally inconsistent with the total dataset.  We exclude these days
from the analysis, which has a small ($\ll 1 \sigma$) effect on our
results.  The visibilities for all other days were consistent with the
estimate of the visibilities from the rest of the data.  A typical
value of the $\chi^2$ of an individual day as compared with the rest
of the dataset is $\chi^2/\nu = 0.986$ with $\nu = 1617$.  For such a
day the probability of exceeding this value under the null hypothesis
is 65\%.

The data on each configuration of each field was also subdivided in
half by time, subtracted, and then compared in this manner.  The
$\chi^2/\nu$ values for these tests on the 08h data (3
configurations), and the 14h, and 20h data (each one configuration)
were $0.990$, $0.993$, $0.967$, $1.015$, and $0.986$ for $\nu=2310$,
$2710$, $3948$, $2310$, and $4848$, respectively.  These results
indicate that our estimate of the thermal noise variance is correct to
within $\sim 2\%$.

\subsubsection{Kolmogorov-Smirnov Tests}

We have used the Kolmogorov-Smirnov (KS) test to determine if the
visibility data are consistent with a Gaussian distribution. This is
the expectation when the signal-to-noise-ratio per baseline-channel
combination is low, as is the case for the $\ell > 1000$ data on any
given field once the bright point sources have been subtracted.  The
quantity of interest is the ratio of the visibility to its estimated
uncertainty
\begin{equation}
t = \frac{V}{\sigma_V}
\end{equation}
If the $\sigma_V$ were known {\it a priori} then $t$ would have a
Gaussian distribution with unit dispersion.  As discussed in the
Appendix, however, this is not the case, and a $4\%$ correction to the
expected standard deviation ($\sigma_t$) of $t$ is
required. Furthermore, as discussed in \S~\ref{subsubsec:chi2}, we
have only determined that our measurements of the thermal noise are
accurate at the $1\%$ level. We therefore apply the KS test to the
data under the null hypothesis that the observed distribution of $t$
is consistent with a Gaussian distribution of zero mean and $\sigma_t
= 1.04$, but vary $\sigma_t$ by $\pm 1\%$ for each dataset.  We find
good consistency with the Gaussian distribution, so to the accuracy
that the noise level has been well-quantified we do not detect
significant non-Gaussianity in the visibility data.

\subsubsection{Power Spectra}
\label{subsubsec:powerspectra}

Two sets of bandpowers $q_{1,B}$ and $q_{2,B}$ (i.e., two power
spectrum estimates) can be compared for consistency by forming
\begin{equation}
\chi^2 = (q_{1,B'} - q_{2,B'})^T \, (F^{-1}_{1,BB'} + F^{-1}_{2,BB'})^{-1}
\, (q_{1,B} - q_{2,B}).
\end{equation}
Under the hypothesis that the likelihood function is Gaussian and the
$q_{i,B}$ are drawn from the same underlying power spectrum with noise
correlations specified by the elements of the inverse of the Fisher
matrices $F^{-1}_{i,BB'}$, this is distributed as $\chi^2$ with $\nu$
degrees of freedom, where $\nu$ is the number of free parameters in
{\it each} spectrum estimate (7).  The $\chi^2/\nu$ values (and
significances) for the 08h/14h, 08h/20h, and 14h/20h power spectrum
comparison are, respectively, $1.40$ ($20 \%$), $1.03$ ($40\%$),
$1.08$ ($38\%$).

\subsubsection{Undifferenced and Doubly-Differenced Power Spectra}

In order to estimate the level of non-celestial signals which we are
removing from the data, we have extracted the power spectra of the
undifferenced data ({\it i.e.}, no ${\it lead}-{\it trail}$ based
ground subtraction).  The power spectrum estimates derived from the
shortest baselines show the effect of ground-spillover, but at $\ell >
600$ they are within a factor of $\sim 4$ (in $\mu$K) of the
differenced spectrum, and are seen to fall strongly with increasing
$\ell$.  Since the CMB signals on the short baselines are up to a
factor of $\sim 10^3$ (in $\mu$K) less than the ground signal on these
baselines, the low power levels in the undifferenced spectrum indicate
that the signal averages down when data from many days and position
angles are combined.  At multipoles $\ell > 1000$ the undifferenced
signal is within $\sim 1 \sigma$ of the differenced spectrum except in
a single bin.  In this bin ($2388 < \ell < 3000$) the undifferenced
power level was originally found to be about twice the thermal noise
level.  This was traced to correlator offsets in a single channel on
several baselines.  Since the $\chi^2$ of the differenced data on
these baselines are consistent within the noise with data from other
baselines and the power spectrum is not significantly affected by the
exclusion of the channel in question, we infer that the contamination
is effectively removed by our differencing procedure.

We have also divided the data from the {\it lead} and {\it trail}
pointings for each field in half, subtracted these, and derived the
power spectrum of the combination of the doubly-differenced data sets
for the three fields.  This is consistent with zero --- $\chi^2/\nu =
1.06 $ for $\nu=7$ DOF (7 bins).  

\subsubsection{Simulated Datasets}

We carried out a series of Monte Carlo tests on the data.  In these
tests, the individual days that went into the deep field datasets for
which we have high sensitivity, 08h and 20h, were resampled with the
observed $(u,v)$ coverage and noise assuming a known input power
spectrum.  Since the signal to noise ratio beyond $\ell \sim 1000$ in
a single day is very low, we extracted only the first two bandpowers
from each day and its corresponding simulated dataset.  We find that
the distribution of the bandpowers of the real data is consistent with
that of the simulated data. The distribution of the rms powers of the
real data within the primary beam area on the dirty maps is likewise
consistent with that of the simulated data.

\begin{figure}
\plotone{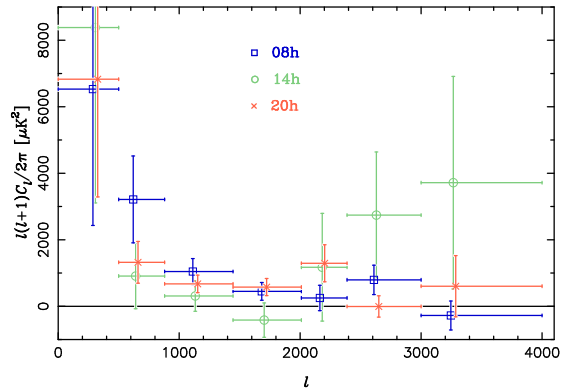}
\caption{Power spectra from the individual CBI deep fields shown for
  the primary binning only.  The 08h, 14h, and 20h data are denoted by
  squares (blue), circles (green), and crosses (red), respectively.  The
  horizontal location of each point on the power spectrum is
  $\ell_{\rm eff}$ and the horizontal error bars indicate the full extent
  of each band.}
\label{fig:deepspec}
\end{figure}

\begin{figure}
\plotone{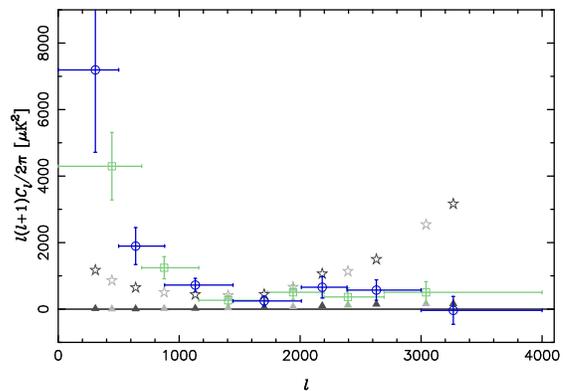}
\caption{Joint CBI deep field power spectra for the primary (blue open
  circles) and alternate (green open squares) binnings.  The placement
  of the points in $\ell$ is determined as in
  figure~\ref{fig:deepspec}.  Also shown are the thermal noise power
  spectrum (open stars) and the residual source power spectrum
  (solid triangles).}
\label{fig:deepspec2}
\end{figure}

\begin{deluxetable*}{llllll}
\tabletypesize{\scriptsize}
\tablewidth{0pt}
\tablecolumns{6}
\tablecaption{Deep Field Bandpower Data $({\rm \mu K^2})$\label{tbl:bandpowers}}
\tablehead{\colhead{Bin}&
\colhead{$\ell_{\rm eff}$}&
\colhead{Joint}&
\colhead{deep 08h}&
\colhead{deep 14h}&
\colhead{deep 20h}}
\startdata
$ 0 -500$     & 307  &  $7193 \pm 2478$ &  $6531 \pm4100$&$8381\pm5274$&         $6829 \pm 3540$\\
$ 500 - 880$  & 640  &  $1896 \pm 556$  &  $3213 \pm 1306$&  $910 \pm983  $& 	 $1322 \pm 627$ \\
$880 - 1445$  & 1133 &  $724 \pm 200$   &  $1045 \pm 389 $&  $310 \pm461  $& 	 $674 \pm 264  $\\
$1445 - 2010$ & 1703 &  $245 \pm 149$   &  $449 \pm 266  $&  $-415 \pm516 $& 	 $578 \pm 257  $\\
$2010 - 2388$ & 2183 &  $657 \pm 323$   &  $251 \pm 382  $&  $1174 \pm1620$& 	 $1293 \pm 557 $\\
$2388 - 3000$ & 2630 &  $573 \pm 308$   &  $795 \pm 442  $&  $2742 \pm1897$& 	 $-5 \pm 320   $\\
$3000 -4000$  & 3266 &  $-37 \pm418$    &  $-276 \pm 435 $&  $3717 \pm3195$& 	 $604 \pm 920  $\\
\enddata
\end{deluxetable*}

\section{Results}
\label{sec:results}

The power spectra of the individual deep fields are shown in
Figure~\ref{fig:deepspec}. These illustrate the field-to-field
consistency of power spectra discussed in
\S~\ref{subsubsec:powerspectra}.  Figure~\ref{fig:deepspec2} shows the
results for the joint fields, as well as the thermal noise and
residual source power spectra.  It is evident that the residual source
correction is small compared to the observed power levels, and that
errors in the thermal noise level power spectrum at the $1\%$ level
indicated in \S~\ref{subsubsec:chi2} are also small compared to
observed power levels over the entire $\ell$-range.  The bandpower
results are summarized in Table~\ref{tbl:bandpowers}.  In this Table,
$\ell_{\rm eff}$ denotes the centroid of the bandpower window
function.  The error bars in this table are $1\sigma$

\begin{figure*}
\epsscale{0.80}
\plotone{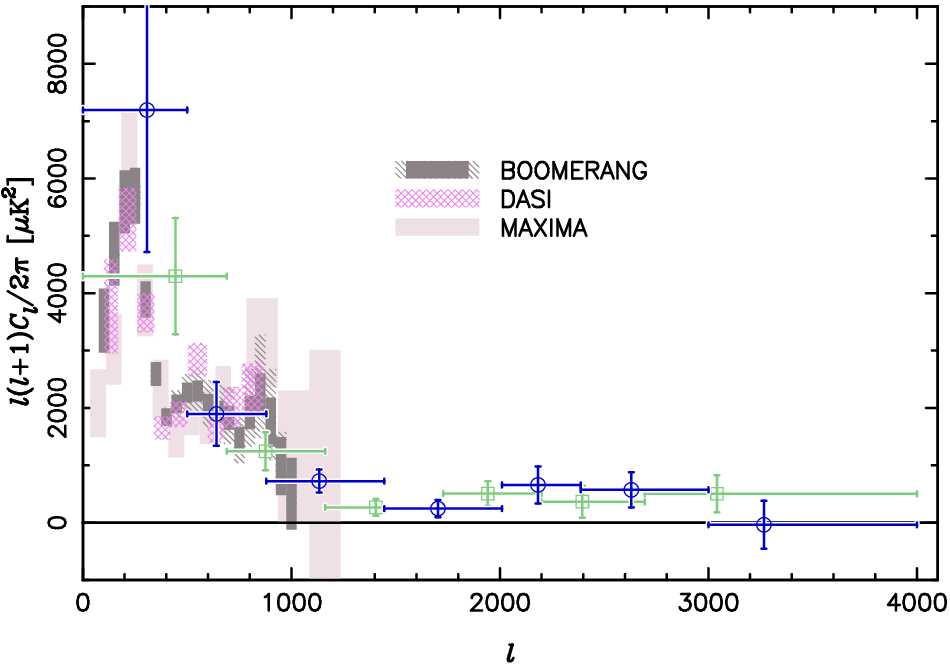}
\caption{Comparison of the CBI deep field power spectrum with
BOOMERANG \citep{newboom}, DASI \citep{dasispectrum}, and MAXIMA
\citep{new_maxima} results. The rectangles indicate the 68\% confidence
intervals on band-power; for BOOMERANG, the solid rectangles indicate
the 68\% confidence interval for the statistical and sample variance
errors, while the hatched rectangles shows the amount by which a
$\pm1\sigma$ error in the beamwidth ($12\farcm9 \pm 1\farcm4$) would
shift the estimates (all up or all down together).  Results from the
CBI primary binning are shown as blue open circles, and the alternate
binning results are shown as green open squares.  The high value of
the power seen in the first bin of the primary and alternate binnings
relative to other observations has rather low significance ($\sim 1.7
\sigma$) due to the large sample variance in the deep field spectra at
low $\ell$; this is discussed further in the text. }
\label{fig:deepspec3}
\end{figure*}

Figure~\ref{fig:deepspec3} compares the CBI deep field results
directly with the BOOMERANG, DASI, and MAXIMA data.  We see that there
is good agreement between all the observations in the range $\ell \sim
600 \rightarrow 1000$, but that the CBI deep field observations are
somewhat higher than the other experiments in the range $\ell<600$.
This discrepancy--- which is not statistically significant--- is
discussed further below and in \S~\ref{sec:conclusion}.  We see in
this figure that the power level drops significantly in the range
$\ell \sim 500\rightarrow 2000$, confirming the results of Paper I,
which were based on a subset of these data, and extending the region in
$\ell$-space over which this drop occurs to higher $\ell$.  Above
$\ell \sim 2000$ the level of power is flat and significantly greater
than zero, contrary to what is expected from standard models of the
intrinsic anisotropy.  We return to this point in \S5.2.

\begin{figure*}
\epsscale{0.80}
\plotone{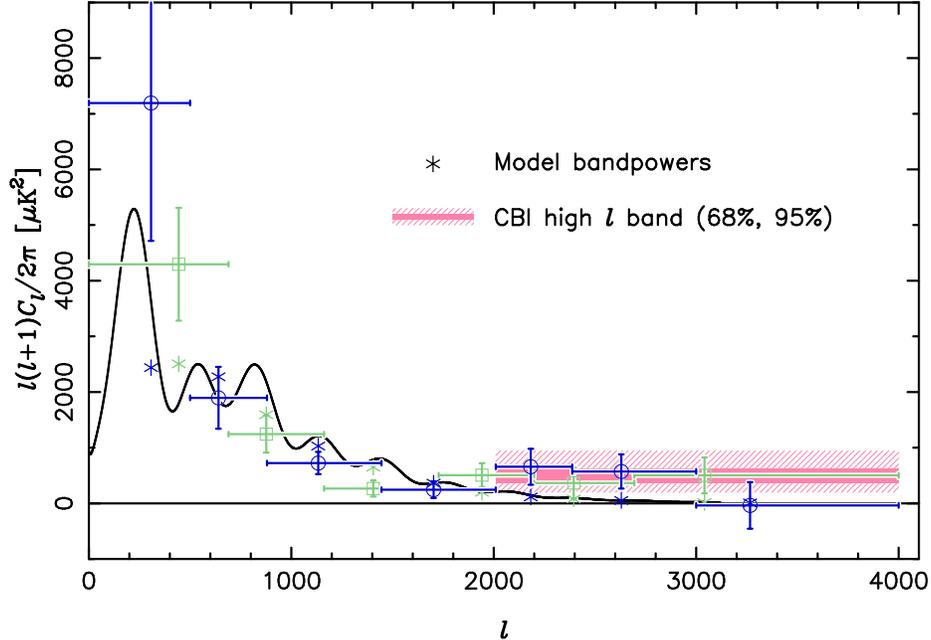}
\caption{The CBI deep field power spectrum: Results from the CBI
primary binning are shown as circles, and the alternate binning
results are shown as squares.  The shaded region shows the 68\% and
95\% central confidence intervals for the high-$\ell$ CBI bin
$2000<\ell<4000$; these was computed by combining the last three
primary bins into a single bin in bandpower and mapping the likelihood
of this band, marginalizing over irrelevant parameters.  The curve
shows the spectrum derived in Paper V as a best fit to ``all-data'' (Boomerang,
CBI, DASI, DMR, Maxima, VSA and earlier observations) using strong priors (age of
the universe $>10$ Gyr, $45 \; {\rm km\,s^{-1}\,Mpc^{-1}} < {H_0}
< 90 \; {\rm km\,s^{-1}\,Mpc^{-1}}$, zero curvature, plus large scale
structure constraints).  Stars show the expected signal in each CBI
band for this model.  The power in the range $\ell \sim 2000
\rightarrow 3500$ differs from the best fit model at the $3.1\sigma$
level.  This excess power is discussed further in the text.  }
\label{fig:deepspec4}
\end{figure*}

In Figure~\ref{fig:deepspec4}, we have plotted the CBI deep spectrum
together with the best-fit model to the CBI mosaic results combined
with the BOOMERANG, DASI, MAXIMA, VSA and earlier observations (``all-data''), derived in
Paper V, subject to the constraints listed in the figure caption. 
Here we also show the expected level of the CBI deep spectrum in
the primary and alternate binnings obtained by integrating this model
over the CBI window functions shown in Figure~\ref{fig:deepwins}.  We
find that the CBI deep spectrum is reasonably consistent with the
theoretical curve at $\ell < 1000$ --- the first two bins of the
standard binning yield $\chi^2/\nu = 1.86$ (84\% significance) to the
theoretical curve, and the first two bins of the alternate binning
yield $\chi^2/\nu=2.03$ ($87\%$ significance).

\begin{figure*}
\plotone{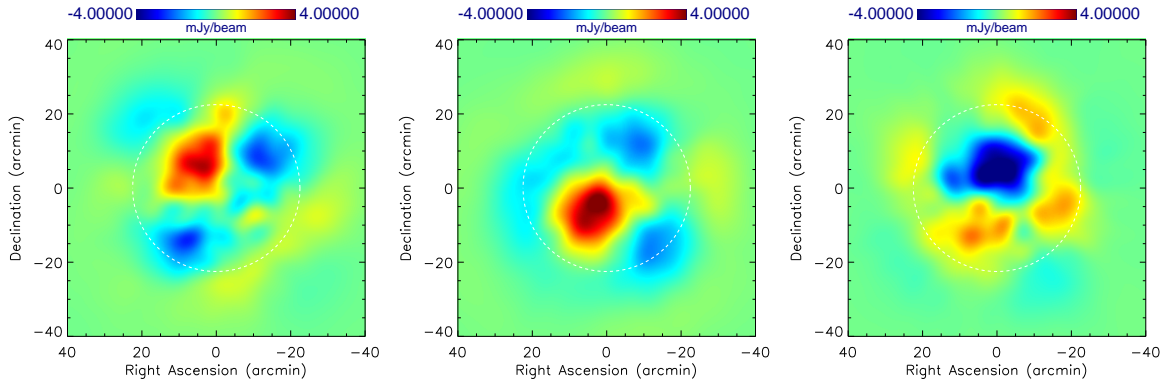}
\caption{Wiener-filtered images of the 08h, 14h and 20h deep fields (see \S~8 of Paper IV).
The positions of the field centers are given in Table 1. The dominant
features in these images have angular scales $\sim 15 \arcmin$, and
are due to structures with multipoles observed on $\sim 1$ meter
baselines, which span the range of the second and third acoustic
peaks.  The approximate fraction of the data contributing to these
features can be seen in Figure~\ref{fig:wtplot}, where the peaks in
the range $500<\ell<800$, correspond to the data contributing to the
dominant features in these images.  Due to the differencing and the
sidelobes of the point spread function, there is some ambiguity in
relating particular features in these raw maps to actual features on
the sky.  Figure 1 of Paper I presents a similar image of the 08h
field, along with the point spread function and one of several
possible deconvolutions.}
\label{fig:images}
\end{figure*}

We have constructed Wiener-filtered images of the sky signal in which
we have subtracted the point source and noise contributions by the
method described in Paper IV and in \citet{bond_and_crittenden}.  The
images for the three fields are shown in Figure~\ref{fig:images}.

\subsection {Limits on Diffuse Foregrounds}

We can use the spectral coverage of the CBI to place limits on
potentially contaminating foregrounds such as Galactic synchrotron
($\alpha \sim -0.7$) and free-free emission ($\alpha \sim 0$). To do this,
we generated a set of 100 realizations of our data set using the
best-fit power spectrum shown in Figure~\ref{fig:deepspec4}, as well
as 100 realizations of a foreground with zero spectral index, typical
of free-free emission, with a flat power spectrum.  We added the two
sets together at various power levels of the foreground, and fitted
the data from $\ell<800$, where we are most sensitive to the spectral
index, to a two-bin ( $\ell<500$ and $\ell>500$) model with a varying
spectral index, finding a single best-fit spectral index using that
model.  Models containing both CMB and foreground components are too
ill-constrained by the data to be useful.  We find that the mean
best-fit spectral index for the zero-foreground simulations is 1.97,
and that the scatter of the individual simulations is 0.34.
Furthermore, we find that the mean best-fit spectral index at
different simulation power levels is well described (1\% error) by
taking the means of the CMB and foreground spectral indices, weighted
by their power levels.
\begin{equation}
\label{eq:jseq}
\alpha_{\rm fit} = \frac{\alpha_{\rm cmb} P_{\rm cmb} + \alpha_{\rm fg} P_{\rm fg}}{P_{\rm cmb}
+P_{\rm fg}}, 
\end{equation}
with $P_{\rm cmb}=2455 \muKt$ and $\alpha_{\rm cmb}=1.97$ for $500<\ell<800$.  The data give a
best-fit spectral index of $1.84\pm 0.34$, $0.4\, \sigma$ from the
zero-foreground value.  The $1\sigma$ upper limit for $\ell<880$ on a
free-free like foreground contribution is $\alpha_{\rm fit}=1.84-0.34 =
1.50$, which, using  Eq~\ref{eq:jseq}, yields $P_{\rm fg} = 744
\muKt$ ($31 \%$ of $P_{\rm cmb}$) for $\alpha_{\rm fg}=0$.  For a synchrotron
spectral index of $-0.7$, we get an upper limit $P_{\rm fg} \leq 521
\muKt$ $(P_{\rm fg}< 21 \%$ of $P_{\rm cmb})$.

\subsection {The Apparent Excess Power at High $\ell$}

An interesting feature of the joint spectrum is the apparent excess
power observed above $\ell=2000$.  To quantify this, we reanalyzed the
data with the last three bins in the primary binning grouped into a
single bin ($2010 < \ell < 4000$).  For the single high-$\ell$ bin we
obtain a best-fit bandpower of $508 \, {\rm \mu K^2}$ with an
uncertainty (from the Fisher matrix) of $168 \muKt$.  In order to more
accurately estimate the uncertainty we have calculated the likelihood
curve of the bandpower in the high-$\ell$ bin explicitly---
marginalizing over the bandpower in the previous bin, which is the
dominant bandpower correlation, and the residual source correction---
and integrated the likelihood function to determine confidence
intervals.  The $68\%$ and $95\%$ central confidence intervals are,
respectively, $(359 \rightarrow 624) \, {\rm \mu K^2}$ and $(199
\rightarrow 946) \, {\rm \mu K^2}$.  Averaged over the high-$\ell$
bin, the model shown in Figure~\ref{fig:deepspec3} predicts $\sim 66
\, {\rm \mu K^2}$ of power.  Power levels this low or lower are
excluded at a significance corresponding to $3.1 \sigma$, and zero
signal is excluded at a significance corresponding to
$3.5\sigma$.  The best-fit values of the other four bins of the
high-$\ell$ analysis are close to those achieved in the primary
(seven-bin) analysis.  The joint deep-field power spectrum is shown
with the high-$\ell$ $95\%$ central confidence interval ($\sim \pm 2
\sigma$) in Figure~\ref{fig:deepspec4}.  These results differ from
what would be obtained from the Fisher matrix error bars since---
although the Fisher matrix calculates the overall curvature
accurately--- the actual likelihood is asymmetric.  Using the Fisher
matrix uncertainties the significance of a detection of non-zero power at $\ell
> 2010$ is $2.8\sigma$, including the contribution of the uncertainty
in the residual source power.

\begin{deluxetable}{ll}
\tabletypesize{\scriptsize}
\tablewidth{0pt}
\tablecolumns{2}
\tablecaption{Effect of Source Treatment on High-$\ell$ Bin\label{tbl:bigbintests}}
\tablehead{\colhead{Source Corrections}&
\colhead{\begin{tabular}{c}Power Level\\ $(\mu {\rm K}^2)\tablenotemark{a}$\end{tabular}}}
\startdata
 No Correction   & $3028^{+850}_{-718}$   \\ 
 OVRO Subtraction Only   & $1074^{+ 449}_{- 361}$ \\ 
 OVRO+NVSS Corrected   & $603^{+372}_{- 279}$    \\ 
 OVRO, NVSS, and Residuals   & $508 \pm 356$   \\ 
\enddata
\tablenotetext{a}{Errors are $\pm 2\sigma$.}
\end{deluxetable}

In order to give a clear idea of the corrections being applied,
Table~\ref{tbl:bigbintests} shows the changes in the high-$\ell$ bin
that result from the application of the individual corrections.
Two-thirds of the source power is eliminated by the OVRO subtraction.
Of the remaining $\sim 500 \, \muKt$ of excess source power, about
80\% is removed by projecting NVSS sources out of the data, and 20\%
is removed by the statistical residual source correction.  As
indicated in \S~\ref{subsubsec:statisticalsources} the known source
projection is robust; the results are insensitive to even large
changes in the projection coefficients, $q$, and to fairly major
corruptions of the covariance matrices such as randomly reassigning
variances between sources.

 The spectral index distribution of the OVRO sources, used in
calculating the residual source correction, covers the range $-1.3 <
\alpha < +0.5$.  However, sources with spectral indices up to
$\alpha=+2$ have been detected \citep[e.g.,][]{gpssample,edgegps}.
We have therefore explored the possibility that a seperate population
of objects, not seen in the NVSS or accounted for in the residual
source correction, might be responsible for the excess.  For this
hypothetical population, we assume a power-law integrated number-flux
density slope $\beta$ at 31 GHz, and we then compare the number
density of sources in this hypothetical population required to produce
the observed excess with the limits from the CBI data, where we found
no sources above the 5-$\sigma$ cutoff in the deep or mosaic fields
which were not correctly identified with NVSS sources, as discussed in
\S~\ref{subsubsec:brightsources}.  Even for the very steep integrated
source count slope of $\beta=-1.5$, which has not been observed at low
flux densities and high frequencies, we find that the number density
of sources in this hypothetical population required to explain the
excess exceeds the upper limit from the CBI observations by
$>3\sigma$.  It is therefore unlikely that such a hypothetical
population of inverted spectrum sources is responsible for the excess.

The apparent excess would be explained if we had underestimated the residual
source correction by a factor of $4.5$.  We have been unable to
construct a model which achieves this while remaining consistent with
the source counts we have derived at 31 GHz, and source statistics at
other frequencies.

\section{Discussion and Conclusions}
\label{sec:conclusion}

In this paper we have presented measurements of the CMB power spectrum
out to $\ell=3500$, beyond the scales probed by BOOMERANG, MAXIMA, and
DASI and well into the damping tail region of the spectrum.  Our
present results confirm our previous detection of a drop in power at
multipoles above $\ell \sim 750$ relative to the level at lower
multipoles, based on a subset of the data used here (Paper I), and
show that the decline in power persists out to $\ell \sim 2000$.  Such
damping is one of the fundamental
predictions of standard cosmological models \citep{silk}.  Below $\ell
\sim 900$ the power levels observed are consistent with those seen in
earlier experiments
\citep{Miller_et_al_1999,boom,leitch_et_al_2000,maxima,dasispectrum,new_maxima}.
At $\ell<500$ the power detected here is greater than that seen in
other experiments including our own mosaic observations (Paper III).
As discussed in \S~\ref{sec:results},  most of this discrepancy is due to
the first bin.  The CBI mosaic spectrum (Paper III), which is based on
a larger area and therefore has lower sample variance than the deep
field results, has power levels at low $\ell$ which show excellent
consistency with the BOOMERANG, DASI, and MAXIMA results. It is
therefore likely that the discrepancy seen in the deep fields is due
to the large sample variance in the low-$\ell$ mode estimates from
these fields.

In Paper V we discuss the constraints which the CBI deep and mosaic
data provide on standard cosmological parameters.  The CBI mosaic
fields are much more powerful for this purpose than the deeps due to
their higher $\ell$-resolution and lower cosmic variance.  The deep
fields, however, provide a robust check on the results: we find
$\Omega_{\rm tot} = 1.09^{+0.11}_{-0.24}$; $n_s = 1.16^{+0.15}_{-0.14}$;
$h=0.61^{+0.10}_{-0.10}$; and an age for the universe of $12.2 \pm 1.8
\, {\rm Gyr}$.  These results assume the weak-$h$ prior (discussed in
Paper V) and use only the deep field power spectra out to $\ell=2000$,
where consistency with standard models of the instrinsic anisotropy is
observed.  More discussion of these results and the analysis method
can be found in Paper V.

Above $\ell \sim 2000$ we detect a $3.1\sigma$ excess in power
relative to the best-fit curve.  Possible explanations of
this excess are:

\begin{enumerate}
\item Data analysis error.  We have tested our analysis by developing
  two independent software pipelines, and by developing a simulation
  program which generates data sets which mimic the $(u,v)$ coverage
  and the distribution of measurement uncertainties of the real data
  set precisely.  We have subjected both real and simulated data
  analyzed with both pipelines to a large battery of tests, and have
  been unable to find significant inconsistencies.
 
\item Instrumental problems.  Any instrumental signals which are stable
 over an 8-minute time span are removed to high precision by the
 differencing; instrumental signals which vary on $<8$ minute time
 spans would be evident in the doubly-differenced power spectra, but
 we see no such signals.  It is also possible that there are
 inadequately modeled instrumental effects.  A prime candidate would
 be pointing errors, but our pointing errors of $\sim 20''$ are too
 small to account for an effect of this magnitude.

\item Primary CMB anisotropy.  This is inconsistent with standard
  theories which fit the low-$\ell$ range well, and is therefore an
  unlikely explanation.

\item Diffuse Galactic foreground.  We cannot rule out this
  possibility with the present observations.  Higher sensitivity
  observations at the same frequency or at a higher frequency could
  test this possibility.  We cannot account for the signal with known
  diffuse foregrounds, and the sensitivity is too low to constrain the
  spectral index of the signal at these multipoles.  However, in view
  of the anomalous component of Galactic emission that has been
  detected at this frequency by \citet{leitch_1997}, we are pursuing
  correlation analysis of the sky images with the {\it IRAS} 100~$\mu$m flux
  density, and with other signals, such as the H$\alpha$
  intensity.

\item Residual point sources. This appeared initially to be a likely
  candidate, but the constraint matrix approach to removing point
  source foregrounds has proved to be remarkably robust, and our 31
  GHz source counts have enabled us to place strong limits on the
  hypothetical population of inverted-spectrum point sources needed to
  produce the excess.  While we cannot strictly rule out such a
  population, a very steep integral counts slope would be implied
  ($\gamma = -1.5$ or steeper), as well as a normalization which is
  inconsistent at the $3\sigma$ level with that determined from the
  CBI-determined source counts.  Future 30 GHz surveys with the GBT
  will allow this issue to be further addressed.

\item Secondary anisotropy. There has been great interest in
predicting the nature of the statistical SZE contribution to the CMB
anisotropy on small angular scales, using both analytical
\citep[e.g.,][]{cooray_nongauss,ma_and_fry} and numerical
\citep[e.g.,][]{bond_and_myers,dasilva,refregier_and_teyssier,seljak_sz,bondtaiwan}
methods.  These works generally predict a crossover between the
intrinsic CMB and SZE signals at $\ell \sim 2000$--$3000$. The level of
SZE anisotropy forecast by theoretical models is in the range of $30
\muKt$ to $300 \muKt$, depending sensitively on the rms mass
fluctuation on large scales in the present universe (characterized by
$\sigma_8$).  Therefore secondary SZE anisotropy is, at some level,
likely to contribute to the excess we report, but it is not clear if
the majority of the observed signal can be attributed to the SZE.  If
SZE anisotropies were to be the cause of the observed excess, values
of $\sigma_8 \gtrsim 1$ would be favored.  For a detailed discussion of
the possible implications of the observed excess for models of SZE
anisotropies, see Paper VI.

\end{enumerate}

Other possible contributors to signals on these angular scales include
the Vishniac effect \citep{vish}, patchy re-ionization
\citep{aghanim96,gruzinovandhu} and gravitational lensing
\citep{blanchardandschneider,coleandefstathiou,seljak96}.  All of
these are expected to be small effects compared to the signal we
observe.  It should be borne in mind that if the signal is due to
non-Gaussian structures then the sample variance errors in our result
and others will have been underestimated; see, for example,
\citet{pengjie_sz} and Paper VI.

\citet{Dawson_et_al} have reported a tentative (1.3$\sigma$) detection
with the BIMA array at the same frequency on a smaller angular scale.
These investigators place a $95\%$ upper limit of $Q_{\rm flat} = 12.4 \,
{\rm \mu K}$, corresponding to $\ell (\ell +1) C_{\ell} /2 \pi = 369
\, {\rm \mu K^2}$ at $\ell_{\rm eff}=5530$. Since this measurement is at
higher $\ell$ than the excess we have found here, it is possible that
these two results are not directly comparable, but as discussed in
Paper VI they could both be affected by secondary SZE anisotropy.

The key result of this paper is the clear demonstration of the
existence of a damping tail to the anisotropy spectrum over the range
$\ell \sim 500 \rightarrow 2000$.  This shows that on average there
are no large deviations of the intrinsic anisotropy spectrum from the
predictions of standard cosmological models over this $\ell$-range.
As discussed in Paper III, these measurements also support the
gravitational instability paradigm for structure formation in the
universe by providing the first direct measurements of the seeds from
which present-day galaxy groups and clusters formed.  In addition we
report a detection of power, significant at the $3.5\sigma$ level, and
$3.1\sigma$ above expected level of intrinsic anisotropy, at $\ell >
2010$.  Higher signal-to-noise ratio and multi-frequency measurements will
be vital in confirming this signal and, if confirmed, determining its
origin. The cosmological implications of the deep field results are
discussed further in Papers V and VI.

\acknowledgements

We thank Roger Blandford, Bryan Butler, Mark Gurwell, Marc
 Kamionkowski, Ken Kellermann, Rick Perley, Sterl Phinney, and Wal Sargent for useful
 discussions.  We acknowledge the invaluable efforts of Russ Keeney,
 Steve Miller, Angel Otarola, Walter Schaal, and John Yamasaki at
 various stages of the project.  We gratefully acknowledge the
 generous support of Maxine and Ronald Linde, Cecil and Sally
 Drinkward, Barbara and Stanley Rawn, Jr., and Fred Kavli, and the
 strong support of the provost and president of the California
 Institute of Technology, the PMA division Chairman, the director of
 the Owens Valley Radio Observatory, and our colleagues in the PMA
 Division.  This work was supported by the National Science Foundation
 under grants AST 94-13935, AST 98-02989, and AST 00-98734.The
 computing facilities at CITA were funded by the Canada Foundation for
 Innovation. LB and JM acknowledge support by FONDECYT Grant
 1010431. SC acknowledges support by CONICYT postdoctoral grant
 3010037. We thank CONICYT for granting permission to operate within
 the Chanjnantor Scientific Preserve in Chile.

\appendix
\section{Noise estimation}

An accurate estimate of the noise in each visibility measurement is
important for power-spectrum estimation.  For the CBI we estimate the
noise as follows. In each 16-min scan (8-min {\it lead} and 8-min {\it
trail}) we form the differences of corresponding {\it lead} and {\it 
trail} integrations. With an integration time of 8.4 s, and allowing 
for slew and calibration time, there are usually about $n=50$ matched 
integrations in one scan. The mean differenced visibility for this  
scan $V_j$ and rms noise $s_j$ is estimated from the individual 
integrations in this scan $V_{ij}$:
\begin{eqnarray} 
   V_j & = & \frac{1}{n} \sum_{i=1}^n V_{ij}  \nonumber \\
   s_j^2 & = & \frac{1}{n-1} \sum_{i=1}^n (V_{ij} - V_j)^2. \nonumber
\end{eqnarray}
The mean and variance are estimated separately for real and imaginary
parts of the visibility and the two variances, which should be equal
if the instrument is working and correctly calibrated, are averaged
together.  In the complete dataset a visibility measurement at a
particular $(u,v)$ point is usually constructed from many ($m$) such
scans taken under different conditions and with different baselines,
so the noise may vary from scan to scan. If the noise on each scan
were known {\it a priori}, the maximum likelihood estimator of the
visibility could be formed by weighted average of the scans
\begin{equation}
   V = \sum_{j=1}^{m} w_j V_j / \sum w_j
\end{equation}
with weights $w_j = 1/s^2_j$, and this estimator would have
a gaussian distribution with variance
\begin{equation}
   \sigma^2 = \frac{1}{ \sum s_j^{-2}}
\end{equation}  
However, this is not the case: when the weights are estimated from the
data, they have their own sampling distribution and the distribution
of estimator $V$ is not gaussian. $V$ (eq.~[A1]) remains an unbiased
estimator of the visibility, but equation (A2) gives a biased
estimator of its variance. The bias depends on the values $n$ and $m$
and on the range of $s_j$ from scan to scan; when data from different
baselines are combined to form a single visibility estimate, the bias
also depends on the relative correlator gains and antenna
temperatures.  We have chosen to use the estimator of equation (A2)
for the variance, but to correct it for the bias. We have found that, for
large $m$ and equal $s_j$, equation (A2) underestimates the variance
by a factor
\begin{displaymath}
f \approx 1 + \frac{4}{n} \approx 1.04.
\end{displaymath}
In simulations we find $f = 1.06$.  We attribute the discrepancy to
fluctuations in the noises and numbers of samples in the actual data,
second-order corrections, and a known and understood $1\%$
overestimate of the noises in our pipeline.  The simulations, and the
$6\%$ noise correction which results from them, take all of these
effects into account. We should therefore increase the variances
computed using equation (A2) by 1.06. We actually used an an earlier,
incorrect, estimate of 1.08 for the factor $f$, so we have slightly
overestimated the noise (by 2\%, comparable with the 2\% uncertainty
in the noise variance discussed in \S~\ref{subsubsec:chi2}). This
overcorrection will have caused a small underestimate of the CMB
band powers.
The cosmological parameter analysis of Paper V corrects
for the effect of this noise misestimation, as will future analyses.

\end{document}